\documentclass[a4paper,12pt]{article}
\usepackage{graphicx}

\begin{document}

\title{{\Large \textbf{Scaling Law for the Distribution of Fluctuations of Share Volume}}}
\author{Taisei Kaizoji\thanks{Corresponding author: E-mail: kaizoji@icu.ac.jp, 
Homepage: http://subsite.icu.ac.jp/people/kaizoji/} and Masahide Nuki \\ 
Division of Social Sciences, \\ International Christian University \\
Osawa, Mitaka, Tokyo, 181-8585, Japan.}
\date{}
\maketitle

\begin{abstract}
We show power-scaling behaviors for fluctuations in share volume, which no other studies have so far done. After analyzing a database of the daily transactions for all securities listed on the Tokyo Stock Exchange, we selected 1050 large companies that each had an unbroken series of daily trading activity from January 1975 to January 2002. We found that the cumulative distributions of daily fluctuations in share volumes can be well described by a power-law decay, and that the cumulative distributions for almost all of the companies can be characterized by an exponent within the stable L$\acute{e}$vy domain $ 0 < \alpha < 2 $. Furthermore, more than 35 percent of the cumulative distributions can be well approximated by Zipffs law, that is, the cumulative distributions have an exponent close to unity. 
\end{abstract}

\newpage

\section{Intoroduction}
There is a long tradition of studies on power-law scaling in financial markets. Scaling of market prices was first reported by Mandelbrot in 1963 in his seminal work on cotton prices \cite{Mandelbrot}. He presented empirical evidence that the cumulative distribution of daily fluctuations in cotton prices followed a power-law asymptotic behavior, characterized by an exponent $ \alpha \approx 1.7 $; he also proposed stable L$\acute{e}$vy distributions \cite{Levy} as candidates for the probability density function of price changes in financial assets. Fama \cite{Fama} analyzed quantitatively the daily data on the share prices of 30 companies over a 5-year period, and showed a power-law distribution of the price changes. Recent studies have confirmed the presence of scaled distributions of price changes in various financial markets: exchange rate markets \cite{Pictet}, German stocks \cite{Lux}, the stock price index \cite{Mantegna}, \cite{Gopi1},and individual stocks \cite{Plerou}. These results suggest the possibility of the existence of universal properties of price changes. \par 
Although trading volume is, like price, an important quantity that characterizes the activities of financial markets, only a few attempts have so far been made to understand the statistical properties of trading volumes\footnote{Gopikrishnan, et.al. \cite{Gopi2} are among those who first studied the statistical properties of share volume. They showed that the distribution of share volume follows a power law with an exponent close to $1.7$.}. In this paper we study the statistical properties of fluctuations of share volume. By analyzing a database of the daily transactions for all securities listed on the Tokyo Stock Exchange, we found that the cumulative distributions of daily fluctuations of share volumes can be well described by a power-law decay and characterized by an exponent within the stable L$\acute{e}$vy domain $ 0 < \alpha < 2 $. Furthermore, we found that the distributions for a number of companies follow Zipffs law, that is, the cumulative distributions of fluctuations of share volume have exponents very close to unity \footnote{Zipffs law is found in various fields, e.g. \cite{Zipf}, \cite{Okuyama} \cite{Aoyama} \cite{Marseli} \cite{Gabaix} \cite{Adamic}.}. \par
The paper is organized as follows. The following section presents the details of our empirical results on the distribution of fluctuations in share volume, and the final section presents some concluding remarks. 

\section{The Distribution of Fluctuations of Share Volume}

We investigated the statistical properties of the daily data of share volume, using a database of the daily volumes for all securities listed on the Tokyo Stock Exchange. We selected 1050 companies that each had an unbroken series of daily trading activity for the entire 27-year period from January 1975 to January 2002. Each time series of daily trading volume had approximately 7000 data points, corresponding to the number of trading days in the 27-year period. \par
The basic quantity studied for individual companies was the volume $ Q(t) $, defined as the number of shares traded in a trading day. 
For each company, we analyzed increments of volume, $ V = Q(t)-Q(t-1) $. Nissin Steel, a large Japanese steelmaker, serves as a typical example of the 1050 companies we analyzed. Fig. 1 shows the probability distribution of the increments in share volume $ V $ for Nissin Steel. At a glance, the contrast with the Gaussian distribution curve is striking. The mean value is close to zero. Skewness is positive, and kurtosis is significantly positive. The important result is that the probability distribution of share volume exhibits apparently greater probability mass in the tails and in the center than dose the standard Normal, which is characterized by the empirical distribution that has been repeatedly observed in various market data\cite{Pagan}.\par
Fig. 2 (a) and (b) show the cumulative probability distributions of increments of share volume $ V $, i.e. the probability for an increment larger or equal to a threshold $ x $, for Nissin Steel in log-log scale. The cumulative distribution $ P(V \geq x) $ shows a power-law decay:

\begin{equation}
    P(V \geq x) \sim \frac{1}{x^{\alpha}}. 
\end{equation}

Regression fits in the region $ 4 \times 10^2 \leq x \leq 2 \times 10^4 $ yield

\begin{equation}
 \alpha = \left\{ 
    \begin{array}{rl}
     0.97 \pm 0.007,  \quad R^2 = 0.996 & \quad\mbox{(positive tail)}\\
     1.02 \pm 0.003,  \quad R^2 = 0.999 & \quad\mbox{(negative tail)}
    \end{array}\right. 
\end{equation}

For both positive and negative tails, a power-law asymptotic behavior well fits the data over the range $ 4 \times 10^2 \leq x \leq 2 \times 10^4 $. These estimates of the exponent $ \alpha $ are well inside the stable L$\acute{e}$vy range, which requires $ 0 < \alpha < 2 $. More noteworthy is that these are very close to unity, i.e., the distribution of share volume for Nissin Steel follows Zipffs law. \par
To confirm the robustness of the above analysis, we repeated this analysis for each time series of share volume for each of the 1050 companies. For all of the companies, the asymptotic behavior of the functional form of the cumulative distributions was consistent with a power law (1). The estimates of $ \alpha $ were sensitive to the bounds of the regression used for fitting. Thus we used the coefficient of determination $ R^2 $ of the linear regression line against a standard and determined the appropriate values of the exponent $ \alpha $. We chose the results of the regression fit as an appropriate value of the exponents where the coefficient of determination was greater than $0.995$. The estimates of the exponent $ \alpha $ were, for all but $6$ of the $1050$ companies, within the stable L$\acute{e}$vy domain, $ 0 \leq \alpha < 2 $. In Figure 3 we show the histogram for $ \alpha $, obtained from power-law regression fits to the positive tails (a) and negative tails (b) of the individual cumulative distributions of all $1050$ companies. Table 1 lists the proportion of observations $ \alpha $ in each class. As shown in Table 1, the most probable value is the class $ 0.9-1.1 $ for the estimates $ \alpha $ of the positive tails and the class $ 1.1-1.3 $ for the estimates $ \alpha $ of the negative tails. Thirty-eight percent of all the estimates $ \alpha $ for the positive tails were between $0.9$ and $1.1$, while $31$ percent of $ \alpha $ for the negative tails were between $0.9$ and $1.1$. 
More than $30$ percent of the companies had a distribution of fluctuations of share volume characterized by an exponent close to unity, ($ 0.9 < \alpha < 1.1 $). 

\begin{table}[hbtp]
\caption{Classification of estimates of the exponent $ \alpha $}
\begin{center}
\begin{tabular}{ccc} \hline
{\it Class}& {\it Positive tail (\%)} &{\it Negative tail (\%)} \\ \hline
$ 0.7 \leq \alpha < 0.9 $ & 9 & 5 \\
$ 0.9 \leq \alpha < 1.1 $ & 38 & 35 \\
$ 1.1 \leq \alpha < 1.3 $ & 32 & 38 \\ 
$ 1.3 \leq \alpha < 1.5 $ & 13 & 14 \\
$ 1.5 \leq \alpha < 1.7 $ & 5 & 5 \\
$ 1.7 \leq \alpha < 1.9 $ & 2 & 2 \\
$ 1.9 \leq \alpha < 2.1 $ & 1 & 1 \\ \hline
\end{tabular}
\end{center}
\end{table}

\section{Concluding Remarks}
We have presented the statistical properties of the distribution of fluctuations in share volume for individual companies. We found (1) that the distribution of fluctuations in share volume for almost all of the companies was consistent with a power-law behavior, characterized by an exponent inside the L$\acute{e}$vy stable regime, $ 0 < \alpha < 2 $. We also found that more than 35 percent of the distributions of fluctuations of share volume followed the Zipffs law approximately.  \par
Share prices are largely determined largely by the relation between supply and demand. Therefore, in order to get a comprehensive understanding of the market dynamics, we need to investigate the relationship between price and volume. Our primitive work \cite{Kaizoji1} on the share returns for the same companies showed that the return distributions were characterized by the double exponential distribution. This result was inconsistent with previous works on stock returns \cite{Gopi1}, \cite{Plerou}. Two questions must be considered next. The first is, What mechanism generates Zipffs law of volume? The second question is whether or not there exist statistical laws that present the relationship between volume and price. \par
Although so far no theory has been established to describe the relationship between price and volume in financial markets\footnote{Our recent work \cite{Kaizoji1} presented a stockastic model of financial markets, which related share returns to trading-volume changes.}, our empirical results give conditions that any empirically accurate theories of stock markets have to satisfy. These empirical and theoretical studies remain for future work. 

\section{Acknowledgment} 
We thank Michiyo Kaizoji for classifying data and encouragements.

\newpage

\begin{figure}[htbp]
\begin{center}
  \includegraphics[width=15cm,height=17cm]{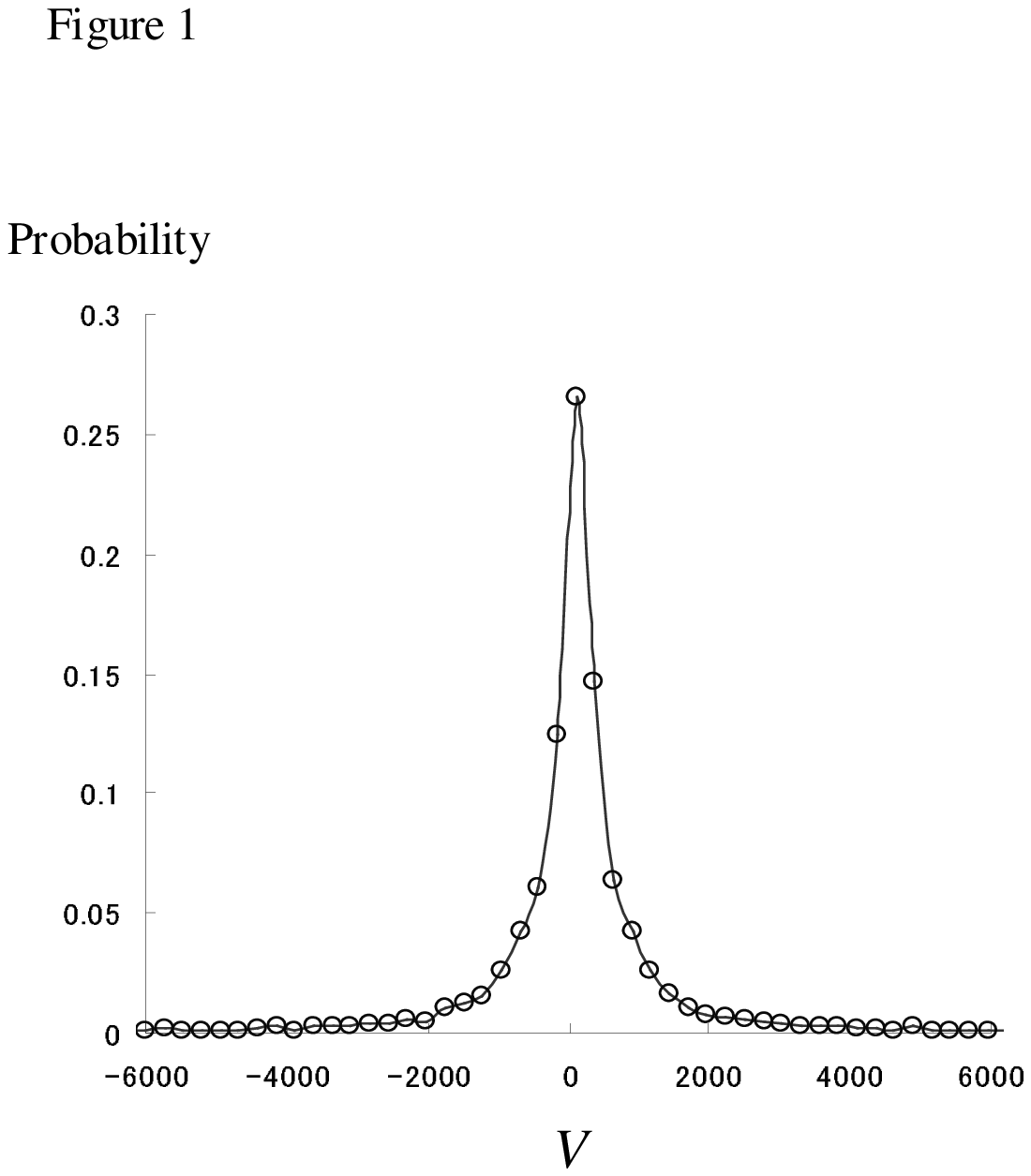}
\end{center}
\caption{Probability density distribution of fluctuations of share volume $ V $ for Nissin Steel for the 27-year period January 1975 to January 2002.  The distribution shows excess kurtosis.}
\label{fig:Figure 1}
\end{figure}

\begin{figure}[htbp]
\begin{center}
\includegraphics[width=15cm,height=17cm]{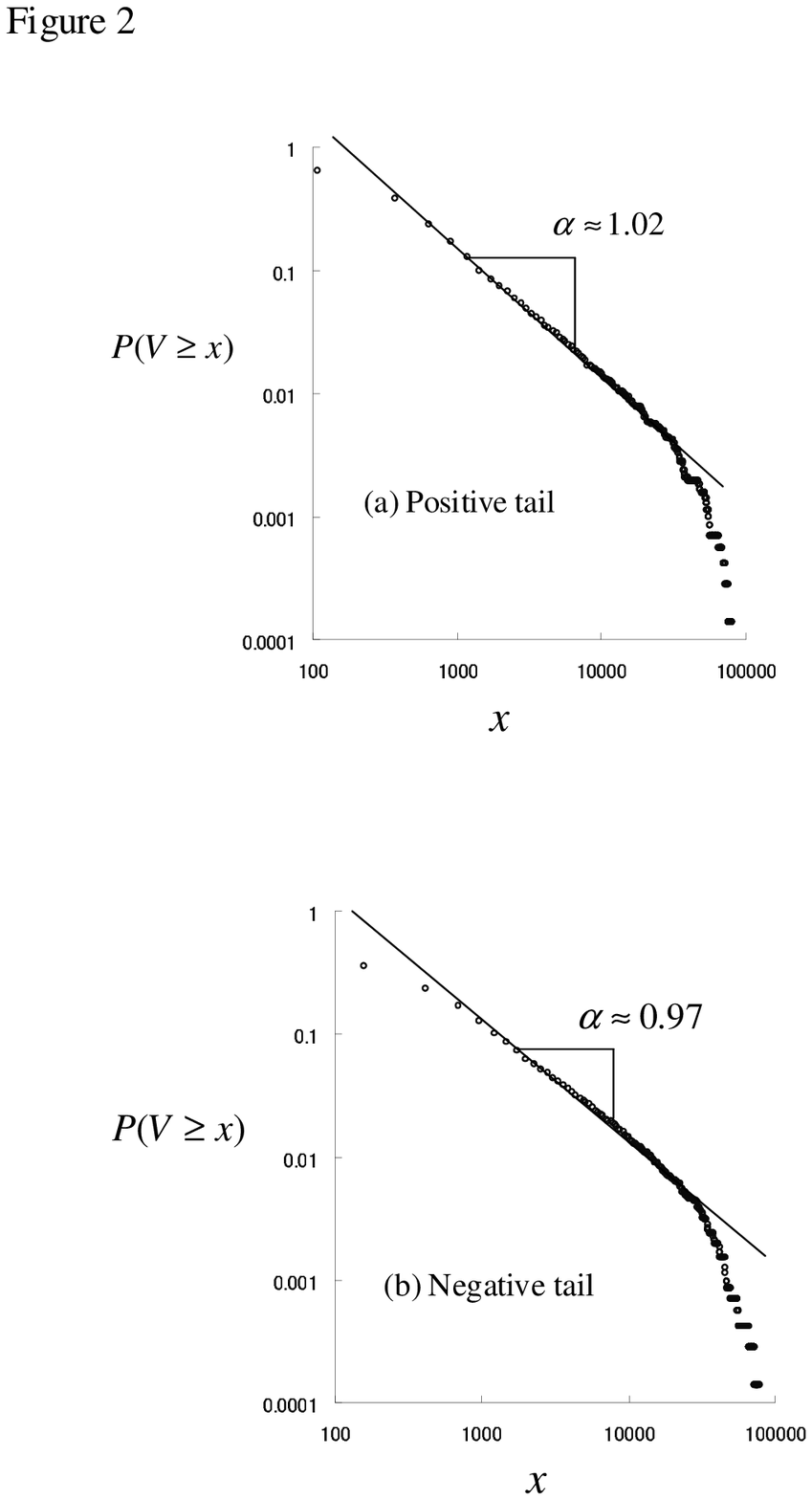}
\end{center}
\caption{Cumulative distributions of (a) positive tails and (b) negative tails of fluctuations of share volume of Nissin Steel for the 27-year period January 1975 to January 2002 in log-log scale. The solid lines are a power-law regression fit in the region $ 4 \times 10^2 \leq x \leq 2 \times 10^4 $. We find $ \alpha = 0.97 \pm 0.007 $ ($ R^2 = 0.996 $) for the positive tail, and $ \alpha = 1.02 \pm 0.03 $ ($ R^2 = 0.999 $) for the negative tail.}
\label{fig:Figure 2}
\end{figure}

\begin{figure}[htbp]
\begin{center}
\includegraphics[width=15cm,height=17cm]{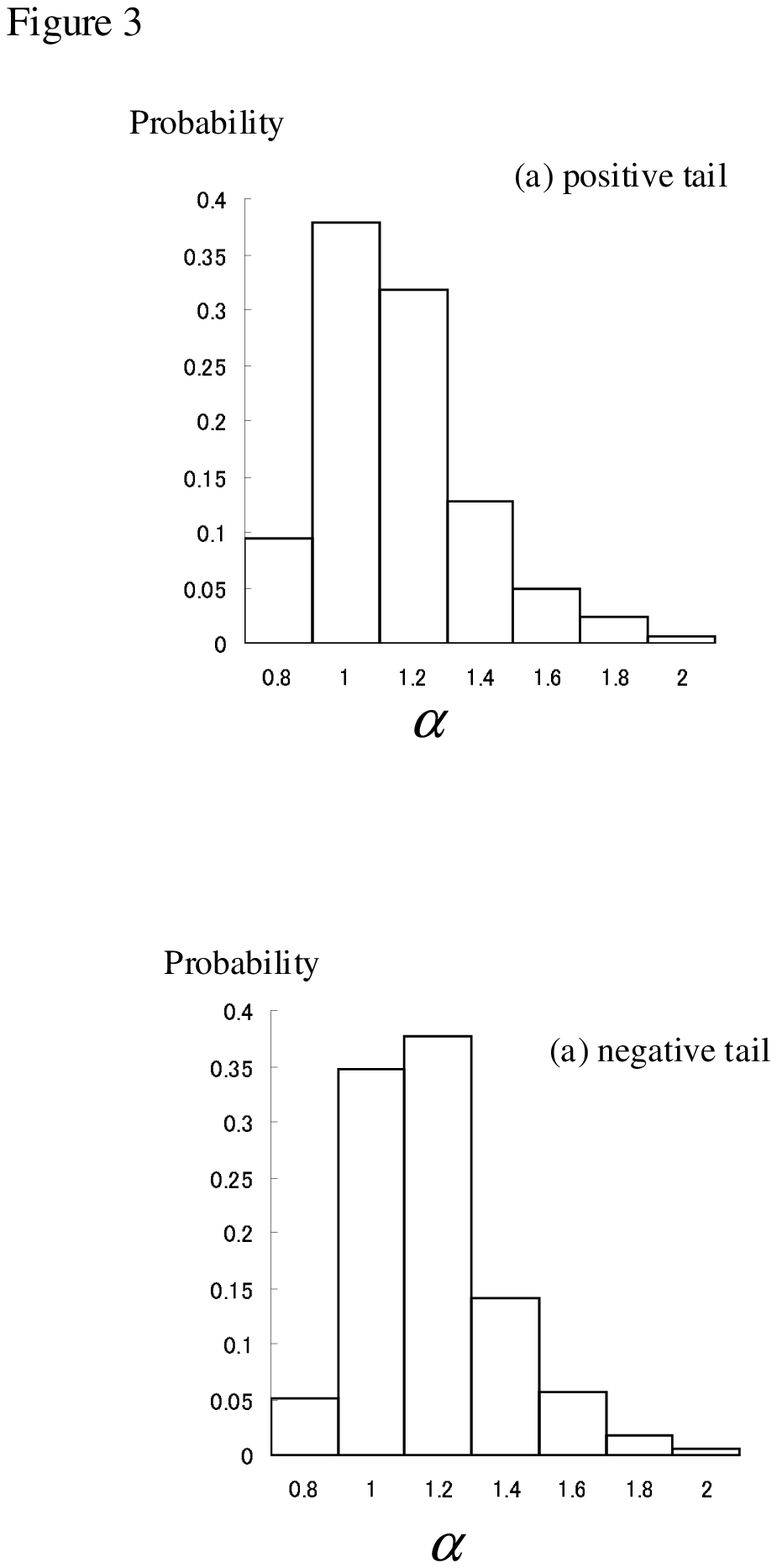}
\end{center}
\caption{Histograms of the exponent $ \alpha $ of (a) positive tails and (b) negative tails of the distributions of fluctuations of share volume. The power-law exponents $ \alpha $ are obtained by a power-law regression fit to the cumulative distribution for each of the 1050 companies, where the coefficient of determination $ R^2 $ of the linear regression line is greater than 0.995.}
\label{fig:Figure 3}
\end{figure}

\end{document}